# Helium-hydrogen immiscibility at high pressures

Yu Wang[1,2], Xiao Zhang[1,2], Shuqing Jiang[1], Zachary M. Geballe[3], Teerachote Pakornchote[3,4], Maddury Somayazulu[3], Vitali B. Prakapenka[5], Eran Greenberg[5], Alexander F. Goncharov[1-3]

[1]*Key Laboratory of Materials Physics and Center for Energy Matter in Extreme Environments, Institute of Solid State Physics, Chinese Academy of Sciences, Hefei, Anhui 230031, China*
[2]*University of Science and Technology of China, Hefei 230026, Anhui, People's Republic of China*
[3]*Geophysical Laboratory, Carnegie Institution of Washington, 5251 Broad Branch Road, Washington, DC 20015, USA*
[4] *Department of Physics, Chulalongkorn University, Bangkok, 10330, Thailand*
[5]*Center for Advanced Radiations Sources, University of Chicago, Chicago, Illinois 60637, USA*

## Abstract

Hydrogen and helium are the most abundant elements in the universe and they constitute the interiors of gas giant planets. Thus, their equations of states, phase, chemical state, and chemical reactivity at extreme conditions are of great interest. Applying Raman spectroscopy, visual observation, and synchrotron X-ray diffraction in diamond anvil cells (DAC), we performed experiments on $H_2$-He 1:1 and $D_2$-He 1:10 compressed gas mixtures up to 100 GPa at 300 K. By comparing with the available data on pure bulk materials, we find no sign of miscibility, chemical reactivity, and new compound formation. This result establishes a new baseline for future investigations of miscibility in the He-$H_2$ system at extreme P-T conditions.

## Introduction

As the two lightest elements, hydrogen and helium are ubiquitous in universe. Moreover, they are the essential elements in the composition of the gas giant planets Jupiter and Saturn in our solar system [1-7]. The existence of helium in the molecular hydrogen ocean and the sedimentation of the helium due to helium-hydrogen separation at high pressures has a significant effect on the evolution of the interiors of gas giants [8-12].

Recently, great progress has been achieved both theoretically and experimentally in understanding the transformation of chemical bonding and formation of novel materials under extreme conditions of high pressure [13-18]. Under pressure, the electronic states of atoms and molecules change, giving elements new chemical character, which can create polymerized states [19, 20], compounds with unusual stoichiometry [13, 21, 22], electrides, and multicenter bonding [23, 24]. It has been found that even chemically inert materials such as rare gas solids made of noble gases become chemically reactive at high pressures [16, 17, 25]. Thus, the chemical inertness of He (which is the second most inert element after Ne) at high pressures[16] cannot be taken for granted. In fact, He forms binary mixed van der Waals compounds with molecular nitrogen - He(N$_2$)$_{11}$ [26] and neon - NeHe$_2$ [27] and it was reported to form an electride compound with Na above 113 GPa [16]. In addition, molecular hydrogen is known to react easily with many elements, forming hydrides with variable composition, including polyhydrides at high pressures [15, 28-31]. Moreover, hydrogen forms binary van der Waals compounds with other diatomic molecules, such as N$_2$ [32, 33]. The latter compounds were found to react to form N-H chemical bonds above 47 GPa [32, 34, 35]. Previous experimental investigations in He-H$_2$ mixtures showed very small miscibility (if any) at room temperature up to 14 GPa[36, 37], which enabled single crystal X-ray diffraction studies of hydrogen to 120 GPa[38]. Theoretical investigations of fluid H$_2$-He mixtures show immiscibility at 100s of GPa, over a temperature range up to at least 3000 K [39-42]. However, to the best of our knowledge there are no predictions about the formation of any compound at low temperatures. In contrast, a strong chemical association, miscibility, and structural changes in He-H$_2$ mixtures were reported recently at pressures as low as 12.6 GPa[43]. This report has been critiqued in a very recent report[44], where the Raman peaks of new phases observed by Lim and Yoo[43] have been attributed to N$_2$ impurity. Here we present Raman spectroscopy and X-ray diffraction (XRD) results in 1:1 H$_2$-He and 1:10 D$_2$-He mixtures up to 100 GPa at 300 K. In contrast to Lim and Yoo [43], we show immiscibility of H$_2$ and D$_2$ with He, as well as the absence of the He-H$_2$ and He-D$_2$ compounds. The complex Raman spectra of He-H$_2$ and N$_2$ doped He reported by Lim and Yoo[43] are due to contamination by N$_2$ and O$_2$, respectively. Our results support the phase diagram reported in Ref. 37 and extend it to higher pressure, assuming that miscibility is not hindered by kinetics.

We performed experiments with the research grade (99.999% purity) 1:1 H$_2$–He and self-produced approximately 1:10 D$_2$–He mixtures made of high purity research grade components. The choice of the compositions was arbitrary as Lim and Yoo[43] reported their effects in a wide range of He-H$_2$ concentrations. The gas mixture was loaded into DACs by compressing up to 0.17 GPa at room temperature. To trap a 1:10 ratio of D$_2$ to He in our high-pressure sample chamber, we pumped D$_2$ to a pressure of 17 MPa and then pumped He to a total pressure of 170 MPa, waited for about 1 hour for the gases to mix, and then sealed the DAC. The connecting gas lines and the high-pressure loading chamber were carefully purged before gas mixing and loading. Raman experiments were performed at ISSP (Hefei, China), GSECARS (APS, Argonne National Lab), and the Geophysical Laboratory (Carnegie Institution of Science) using similar setups which include solid state lasers (473, 488, 532, and 660 nm excitation wavelengths), ultralow low-frequency notch filters (down to 10 cm$^{-1}$ Raman coverage), and wavelength dispersive single grating spectrographs with CCD detectors[16, 34, 45]. X-ray diffraction measurements were performed at the undulator XRD beamline at GSECARS. Typical X-ray beam size in all the experiments was 2-3 μm. Pressure was determined via ruby fluorescence scale[46], Raman of the stressed diamond anvil [47], gold and ruby XRD sensors [48, 49] (the latter one below 50 GPa). In the control XRD experiments on He the sample was laser annealed up to 2000 K at 70 GPa using nanocarbon immersed in He; this did not affect the results substantially indicating that He remains in quasihydrostatic conditions. All the measurements have been performed at room temperature. The figures presenting the proposed phase diagram and the additional XRD and Raman data are in Supplemental Material [50].

After gas loading the $H_2$–He gas mixture to approximately 0.2 GPa, the system is a single miscible fluid (Figs. 1 and S1 [50]). The Raman spectra show the bands characterizing $H_2$ molecules: rotons and vibron corresponding to change in rotational and vibrational molecular states. We find that the spectra are independent of the sample position and the sample looks uniform, indicating miscibility (Fig. 1). The Raman frequency of the vibron mode in a disordered state strongly depends on the $H_2$ concentration and the matrix material [37, 51]. It is shifted to higher frequency compared to pure bulk $H_2$ because of the difference in environment. Upon the transition to a state where $H_2$ solidifies (Fig. S1 [50]), the vibron mode shifts down in frequency abruptly due to a cooperative effect of intermolecular coupling in the ordered state. The sample shows three distinct areas in this regime: $H_2$-rich solid $S_1$, He-rich fluid $F_2$, and a mesoscopic mixed region (Figs. 1, 2); the Raman spectra of hydrogen clearly characterizes all these three areas (Fig. 1).

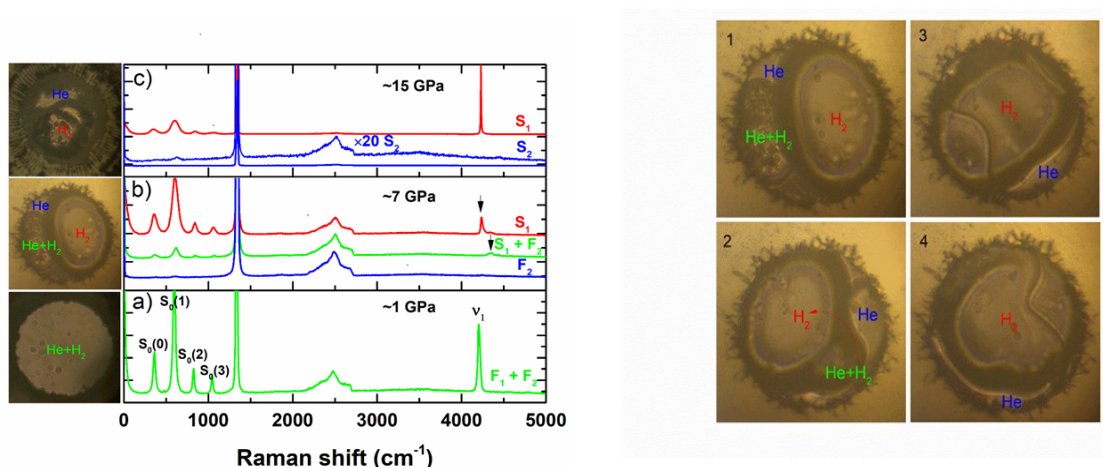

**Figure 1 (left).** Microphotograph and corresponding Raman spectra of $H_2$-He mixtures at different phases. He-rich areas are marked in blue "He". $H_2$–rich areas are marked in red "$H_2$". A mixture He and $H_2$ is marked in green "He + $H_2$". a) Pressure below 5 GPa. Both hydrogen and helium are in a fluid state. b) Pressure above the solidification line of $H_2$–rich fluid in Fig. S1 [50]. Hydrogen is solid and helium is in a fluid state. Corresponding spectra are in same color. c) Pressure above the solidification line of He–rich fluid. Both hydrogen and helium are in solid state. Pressure is inferred from the position of the hydrogen vibron of $F_1+F_2$ and $S_1$ as ruby is absent in this set of experiments to avoid the fluorescence, which can interfere with the Raman detection of weak hydrogen vibron modes.

**Figure 2 (right).** Temporal changes in morphology of $H_2$-He mixtures in the stability range of $S_1$-$F_2$ conglomerate when also exposed to a low-power laser beam (at ~7 GPa). Smaller hydrogen-rich crystals surrounded by fluid helium migrate with time to merge with the bigger hydrogen crystals.

No measurable $H_2$-vibron frequency irregularity is observed through the solidification of He, and the vibron frequency in both $S_1+F_1$ and $S_1+S_2$ states follows that of a pure bulk hydrogen, indicating the formation of a $H_2$-rich solid in our experiments (Fig. 3). We find no sign of the $S_1$' and $S_3$ solids that seemed to be created in the experiments of Ref. 43. In the previous work, phases $S_1$', $S_2$, and $S_3$ were characterized by "novel" Raman bands at 140, 2400, and 3200 cm$^{-1}$, and additional higher frequency $H_2$ vibron modes. The present measurements are sensitive to these Raman signatures, and yet no signal is detected (Fig. 4). Raman experiments in the $D_2$-He system up to 54 GPa reveal a similar behavior and also no extra lines have been recorded (Figs. S2 and S3 [50]). These latter experiments were designed to investigate the isotope effect on the H-He vibron frequency at 2400 cm$^{-1}$ proposed by Lim and Yoo[43], but the experiments yielded no measurable Raman peak at 2400 cm$^{-1}$ or at an isotopically shifted frequency.

Our Raman measurements aimed at the He-rich solid $S_2$ show a small peak at exactly the same frequency as that of the vibron mode of $H_2$- rich solid $S_1$, strongly suggesting that it is originated in the $S_1$ region of the sample chamber where it is the dominant Raman signal (Fig. 4). In both S1 and S2 solids, there are no other Raman bands at 140, 2400, and 3200 cm$^{-1}$, and additional higher frequency $H_2$ vibron modes as reported in Ref. 43 (Figs. 1, 4 and Fig. S3 [50]). Based on the absence of any extra Raman peaks in $S_2$ and the signal to noise ratio exceeding 1000:1 for the $H_2(D_2)$-vibron, we conclude that miscibility of $H_2(D_2)$ in He is less than 0.1% and there is no detectable formation of any chemical bonds between He and H(D)[43]. We note that if a compound of He and $H_2$ formed, it would be an inclusion compound stabilized via an enthalpy gain due to long-range Coulomb interactions [52]. However, we find no sign of an extra lattice mode (Figs. 1, 3, 4), which would likely to appear in this case (cf. Ref. 43).

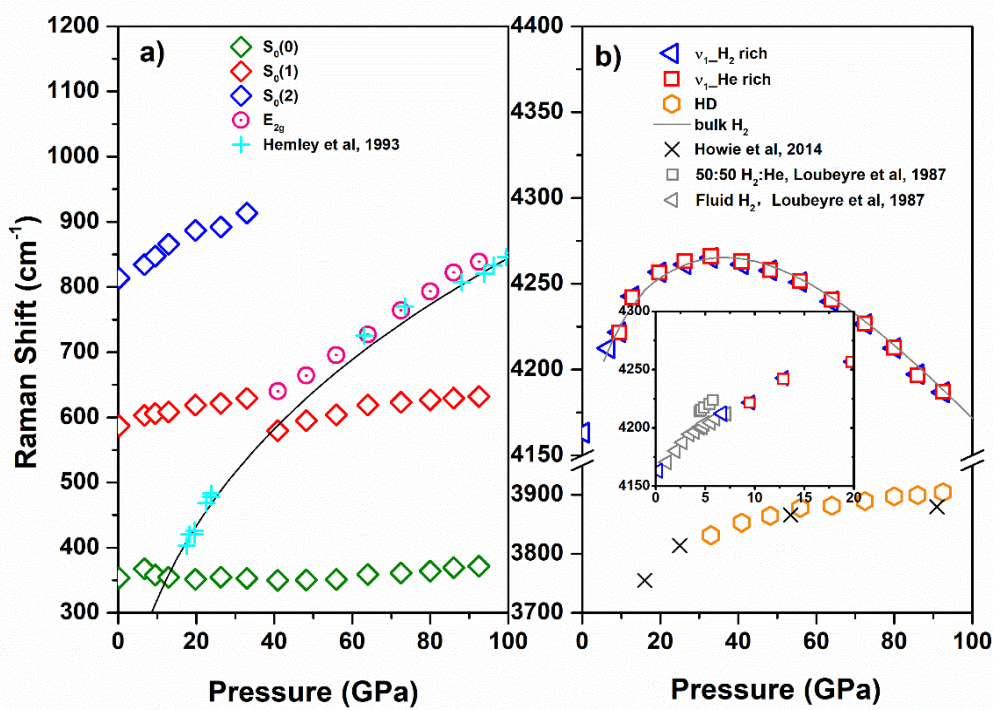

**Figure 3**. The pressure dependence of the Raman frequencies of the $H_2$-rich fluid and solid in He medium as a function of pressure. Black solid and gray lines and light blue crosses[53] are the literature data (see Ref. 54 and References therein) for $H_2$. An inset shows the details of low-pressure behavior, the results of previous measurements are from Ref. 36. Our spectra indicate the presence of HD molecules (<2%) as the major impurity in $H_2$. Crosses in the right panel show the previously measured pressure dependence of the vibron mode of HD molecules which formed in 70:30 $H_2$-$D_2$ gas mixture [55].

The low-frequency Raman spectra of $H_2$-rich solid $S_1$ probe the quantum rotational transitions (rotons), which are weakly pressure dependent, and a translational mode (phonon), which is strongly pressure dependent (Fig. 3). The phonon mode is relatively weak and it can be observed via the rotational-vibrational coupling with the roton modes, which can be clearly viewed as a dip that first appears at the left side of an $S_0(1)$ roton mode at nearly 600 cm$^{-1}$ and then moves with pressure to the right side of this band (Fig. S4 [50]). The pressure dependencies of the Raman modes show an avoided crossing of the phonon roton modes (Fig. 3). These measurements demonstrate that $S_1$ solid is well ordered as otherwise one would expect to see a relaxation of the Raman selection rules. Furthermore, the comparison of the low-frequency spectra show that they are indistinguishable from those of the pure bulk $H_2$ (Fig. S4 [50]). Our experiments in He-$D_2$ system demonstrated qualitatively similar results (Fig. S3 [50]). On the contrary, measurements of these excitations in mixed disordered $H_2$-$D_2$ crystals [55] show a broader spectrum and no coupling of the phonon and roton modes.

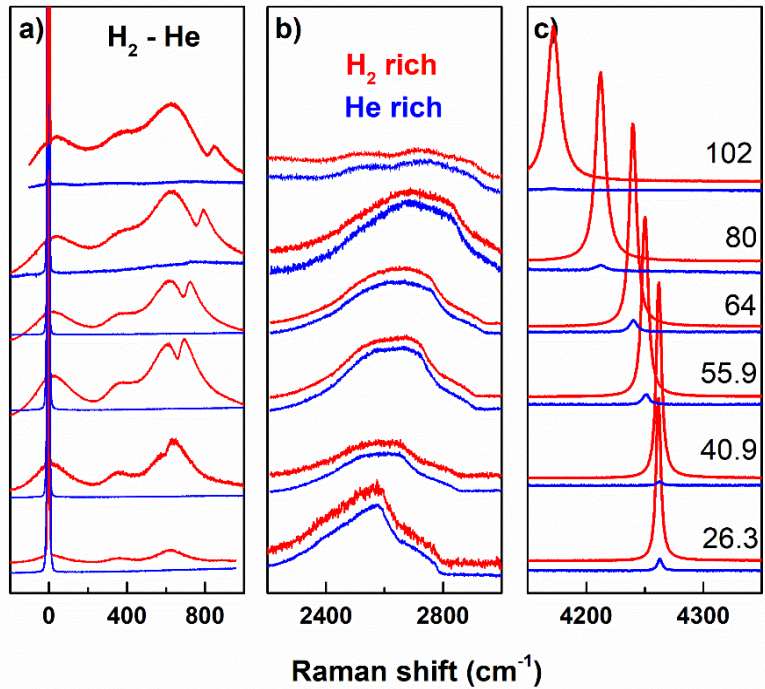

**Figure 4.** Raman spectra of $H_2$-He mixtures at different pressures in the range of rotational-translational (a), second-order diamond (b), and vibrational (c) modes. The spectra of hydrogen- ($S_1$) and helium- ($S_2$) rich parts are shown in red and blue respectively. The spectra of $S_2$ solid have a small contribution from the $S_1$ solid, which surrounds it.

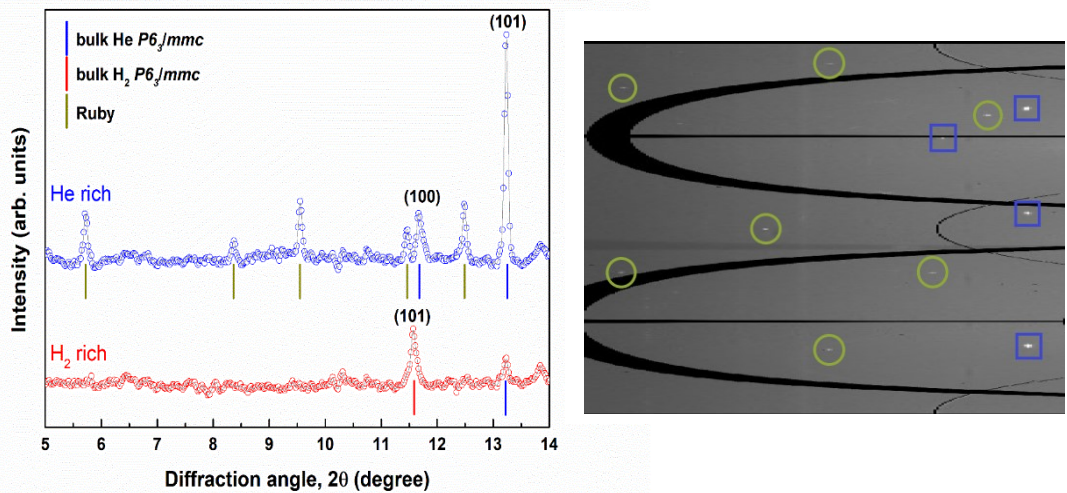

**Figure 5**. X-ray diffraction patterns of $H_2$-He mixtures at 37.5 GPa measured using a wide angle scan to make detectable single crystal reflections (see, for example, Ref. 56). The x-ray wavelength is 0.3344 Å. Blue and red vertical ticks with the peak indexing are presented for the He and $H_2$ respectively. Yellow vertical ticks correspond to the positions of the ruby XRD peaks, which were used to determine the pressure [49]. The right panel is the 2D XRD image (cake) of the He rich solid.

Synchrotron XRD measurements have been performed on the same $H_2$-He sample at 37.5 GPa and 102 GPa (Fig. 5). The DAC has been rotated along the vertical (Ω) axis to match the Bragg conditions in single crystals of He- and $H_2$-rich solids. We have been able to observe a few low-*hkl* diffraction peaks, which identify the lattice symmetry of these crystals (Fig. 5; Table 1). The results show that both solids form an hcp lattice, in agreement with previous reports on He up to 53 GPa [38, 47, 57, 58] and $H_2$ (*e.g.* Ref. 38). These data have been compared to the results of three other experimental XRD runs (without hydrogen) up to 74 GPa, where He was the sample or served as a transmitting medium, and the lattice parameters were determined similarly to the described above (Fig. 6). The (101) XRD peak of the $H_2$ rich solid is documented in our $H_2$-He experiment at 37.5 GPa, and its position is consistent with hcp $H_2$ reported previously (*e.g.*, Ref. 38).

The lattice parameters and the unit cell volume of He measured here are in a good agreement with those measured previously (Fig. 6), in which pure He was used as a sample [57, 58]. A small disagreement may be related to systematic errors due to the use of the energy dispersive method in these works. Our results for He-$H_2$ system and for pure bulk He are in a good agreement with each other (Fig. 6). This indicates that He-rich solid $S_2$ prepared in mixtures of $H_2$ and He can accommodate very little $H_2$ impurity (if any). The upper limit of the $H_2$ doping content can be estimated using Vegard's law based on the difference in the lattice parameters of $H_2$ and He which both form hcp lattices in the explored pressure range. For example, at 100 GPa the difference in the lattice parameters is about 10%. The lattice parameters of He solid determined in this work are accurate to about 0.1%. Thus, the maximum amount of $H_2$ admixture in He rich solid determined from our XRD experiments is about 1%.

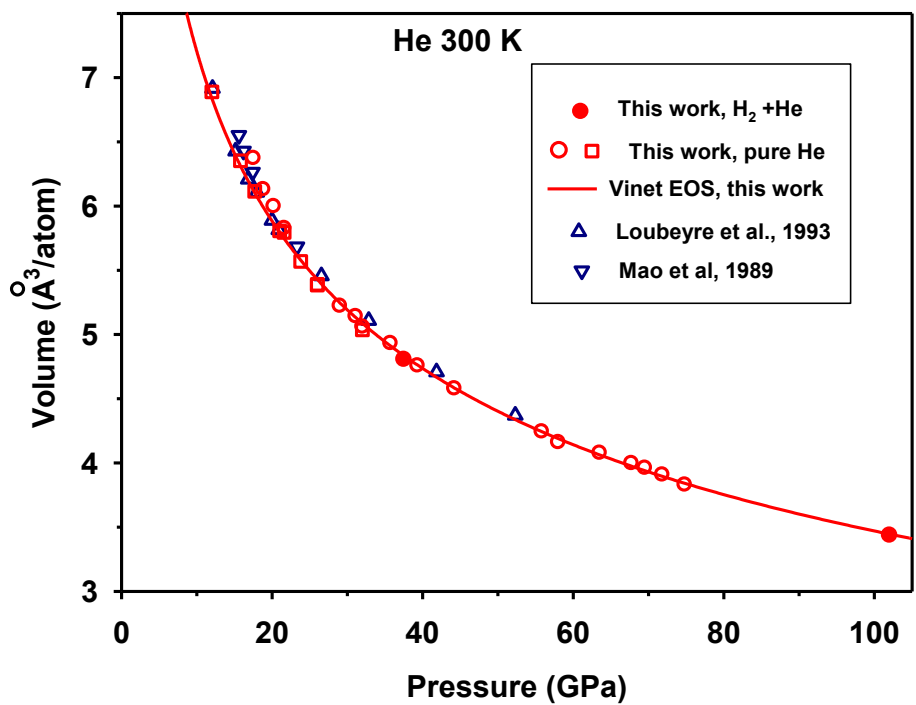

**Figure 6**. The pressure-volume equation of state of solid He measured up to 75 GPa. The results of this work measured both in pure He (squares) and He-rich solid in He-H$_2$ mixture (circles) are compared to those previously determined in pure He [57, 58]. The results of this work are fit with the Vinet equation of state with a fixed $V_0$=31.08 Å$^3$/atom [59], yielding the following parameters: $K_0$=0.039(2) GPa, $K_0$'=8.63(7).

Our Raman and XRD experiments in He-H$_2$ mixtures up to 102 GPa show that H$_2$ and He form the corresponding solid phases. Each phase does not reveal any measurable difference in structural, thermodynamic, or vibrational properties to those of pure materials. Providing that hydrogen and helium are sufficiently mobile at high pressures (*e.g.* Refs. 60-62) and the observed immiscibility is not the kinetic effect, this finding extends the main result of Ref. 36 to higher pressure, strongly suggesting that there is no measurable miscibility of H$_2$ and He in solid phases. How can these results be reconciled with the report about the additional Raman bands at 140, 2400, 3200 cm$^{-1}$, and above 4280 cm$^{-1}$ in the He-rich solid [43]?

First of all, we note that these new bands closely correlate with those reported in the Raman spectra of the H$_2$-N$_2$ mixed system reported previously[32, 34, 35]. These bands can be naturally assigned to the excitations of the H$_2$-N$_2$ van der Waals crystals. In this regard, the disappearance or substantial diminishing of the 2400 cm$^{-1}$ (N$_2$ vibrons) and 4200 cm-1 (H$_2$ vibrons) bands and appearance of the 3200 cm$^{-1}$ band (N-H stretch) reported in Ref. 43 at 50 GPa is closely correlated with the polymerization of the H$_2$-N$_2$ mixed crystal at these conditions [34, 35]. In accord with this, our XRD results indicate that hcp He-rich solid remains stable up to at least 102 GPa similarly to the behavior of pure He suggesting that N$_2$ impurity goes to the H$_2$ rich solid in experiments of Lim and Yoo [43]. As such, we rule out the recently reported chemical association, miscibility and structural change of He and H$_2$ [43]. Similar conclusions have been reached in the independent investigations reported recently [44]. Lim and Yoo [43] point out that the Raman bands in the 2400 cm$^{-1}$ range are somewhat dissimilar in N$_2$ doped He (and also pure bulk N$_2$) and their He-H$_2$ system. However, they compare different systems so these differences can be easily reconciled. Indeed, N$_2$ and He at high pressures form a (N$_2$)$_{11}$He crystal [63], the N$_2$-vibron spectra of which are somewhat different from pure bulk N$_2$. Furthermore, we argue that N$_2$ impurities in Lim and Yoo's He-H$_2$ experiments go into the H$_2$ crystal, not the He crystal, so their control experiment in the N$_2$-He system is not representative. In contrast to the N$_2$-He system, the N$_2$ vibron spectra in the H$_2$-N$_2$ system [34, 44] are very similar to those reported by Lim and Yoo's in terms of a number of the bands and their frequencies. Please note that the N$_2$ vibron spectra of the H$_2$-N$_2$ system depend on the composition, so a small disagreement is possible as the H$_2$ content is expected to be larger in the experiments of Lim and Yoo[43] compared to studies in which N$_2$ was loaded intentionally[32, 34, 35, 44]. Moreover, one should take into account the difference in the spectra of bulk solids (*e.g.*, bulk pure N$_2$) and N$_2$ impurities in H$_2$. In the pure bulk materials there are collective effects, which results in splitting of the vibrons and their shifts due to intermolecular coupling. On the contrary, the impurity molecules, which are uncoupled, will have different vibron frequencies. This has been carefully documented for $H_2$ vibron impurity modes by Loubeyre et al. [51].

Second, Lim and Yoo[43] performed a control experiment in the $N_2$-He experiment and found yet another set of extra Raman lines, which cannot be attributed to nitrogen molecules. Therefore, they argued about the $N_2$ contamination in their He-$H_2$ system. We have compared their frequencies in the $N_2$-He system at different pressures with the literature data on the pressure dependence of the Raman frequencies in $O_2$ [64, 65]. The agreement is almost perfect (Fig. S5 [50]) making clear that Lim and Yoo reported the behavior of $O_2$ doped $N_2$-He system, not a pure $N_2$-He system. Our data and the data of Turnbull et al. [44] do not show the 1600 cm$^{-1}$ peak and others at lower frequencies because these works studied pure systems that do not contain $N_2$ and $O_2$

In conclusion, our combined Raman and XRD experiments on $H_2$($D_2$)-He system up to 102 GPa at 300 K show the occurrence of only two solids, $H_2$ and He. We constrain the impurity concentration in these solids to be below 1 %, thus revealing a very large miscibility gap at the explored conditions of pressure and temperature. Our experiments explain the extra Raman bands reported recently by Lim and Yoo [43] as due to the presence of $N_2$ and $O_2$ impurities in their samples. As such, we rule out the recently reported chemical association, miscibility and structural change of He and $H_2$ [43].

This work was supported by the NSF EAR-1763287, the Army Research Office, the Deep Carbon Observatory, the Carnegie Institution of Washington, the National Natural Science Foundation of China (grant numbers 11504382, 21473211, 11674330, and 51727806), the Chinese Academy of Science (grant number YZ201524), and a Science Challenge Project No. TZ201601. A.F.G. was partially supported by Chinese Academy of Sciences Visiting Professorship for Senior International Scientists Grant 2011T2J20 and Recruitment Program of Foreign Experts. A portion of this work was performed at GeoSoilEnviroCARS (The University of Chicago, Sector 13), Advanced Photon Source (APS), Argonne National Laboratory. GeoSoilEnviroCARS is supported by the National Science Foundation - Earth Sciences (EAR - 1634415) and Department of Energy- GeoSciences (DE-FG02-94ER14466). This research used resources of the Advanced Photon Source, a U.S. Department of Energy (DOE) Office of Science User Facility operated for the DOE Office of Science by Argonne National Laboratory under Contract No. DE-AC02-06CH11357.

**Table 1. The d-spacings of He- and $H_2$ –rich solids (λ=0.3344 Å) observed at 37.5 GPa using a wide scan single-crystal XRD measurements (Fig. 5). To determine the lattice parameters of $H_2$, we used the literature data for the c/a ratio [38].**

|       | *hkl* | *d*   | *2θ*   | *a*      | *c*      |
|-------|-------|-------|--------|----------|----------|
| He    | 101   | 1.449 | 13.250 | 1.894(2) | 3.093(3) |
| He    | 100   | 1.640 | 11.700 |          |          |
| $H_2$ | 101   | 1.656 | 11.587 | 2.177(4) | 3.495(6) |

**Bibliography.**

1. W. B. Hubbard, Science **214** (4517), 145-149 (1981).
2. D. J. Stevenson, Annual Review of Earth and Planetary Sciences **10**, 257-295 (1982).
3. T. Guillot, Science **286** (5437), 72-77 (1999).
4. U. vonZahn and D. M. Hunten, Science **272** (5263), 849-851 (1996).
5. U. von Zahn, D. M. Hunten and G. Lehmacher, Journal of Geophysical Research-Planets **103** (E10), 22815-22829 (1998).
6. B. J. Conrath, D. Gautier, R. A. Hanel and J. S. Hornstein, Astrophysical Journal **282** (2), 807-815 (1984).
7. T. Guillot, Planetary and Space Science **47** (10-11), 1183-1200 (1999).
8. W. B. Hubbard, Astrophysical Journal **162** (2), 687-& (1970).
9. D. J. Stevenson and E. E. Salpeter, Astrophysical Journal Supplement Series **35** (2), 221-237 (1977).
10. J. M. McMahon, M. A. Morales, C. Pierleoni and D. M. Ceperley, Reviews of Modern Physics **84** (4), 1607-1653 (2012).
11. J. E. Klepeis, K. J. Schafer, T. W. Barbee and M. Ross, Science **254** (5034), 986-989 (1991).
12. D. J. Stevenson, Journal of Physics F-Metal Physics **9** (5), 791-801 (1979).
13. W. Zhang, A. R. Oganov, A. F. Goncharov, Q. Zhu, S. E. Boulfelfel, A. O. Lyakhov, E. Stavrou, M. Somayazulu, V. B. Prakapenka and Z. Konôpková, Science **342** (6165), 1502-1505 (2013).
14. Y. M. Ma, M. Eremets, A. R. Oganov, Y. Xie, I. Trojan, S. Medvedev, A. O. Lyakhov, M. Valle and V. Prakapenka, Nature **458** (7235), 182-185 (2009).
15. A. P. Drozdov, M. I. Eremets, I. A. Troyan, V. Ksenofontov and S. I. Shylin, Nature **525** (7567), 73-76 (2015).
16. X. Dong, A. R. Oganov, A. F. Goncharov, E. Stavrou, S. Lobanov, G. Saleh, G.-R. Qian, Q. Zhu, C. Gatti, V. L. Deringer, R. Dronskowski, X.-F. Zhou, V. B. Prakapenka, Z. Konôpková, I. A. Popov, A. I. Boldyrev and H.-T. Wang, Nat Chem **9** (5), 440-445 (2017).
17. E. Stavrou, Y. Yao, A. F. Goncharov, S. S. Lobanov, J. M. Zaug, H. Liu, E. Greenberg and V. B. Prakapenka, Physical Review Letters **120** (9), 096001 (2018).
18. P. M. Celliers, M. Millot, S. Brygoo, R. S. McWilliams, D. E. Fratanduono, J. R. Rygg, A. F. Goncharov, P. Loubeyre, J. H. Eggert, J. L. Peterson, N. B. Meezan, S. Le Pape, G. W. Collins, R. Jeanloz and R. J. Hemley, Science **361** (6403), 677-682 (2018).
19. X. Dong, A. R. Oganov, G. Qian, X.-F. Zhou, Q. Zhu and H.-T. Wang, arXiv:1503.00230 [cond-mat.mtrl-sci] (2015).
20. R. J. Hemley, Ann. Rev. Phys. Chem. **51**, 763-800 (2000).
21. E. Stavrou, S. Lobanov, H. Dong, A. R. Oganov, V. B. Prakapenka, Z. Konôpková and A. F. Goncharov, Chemistry of Materials **28** (19), 6925-6933 (2016).
22. A. F. Goncharov, S. S. Lobanov, I. Kruglov, X.-M. Zhao, X.-J. Chen, A. R. Oganov, Z. Konôpková and V. B. Prakapenka, Physical Review B **93** (17), 174105 (2016).
23. W. Grochala, R. Hoffmann, J. Feng and N. W. Ashcroft, Angew. Chem. Int. Ed. **46**,

3620 – 3642 (2007).
24. M.-S. Miao and R. Hoffmann, Accounts of Chemical Research **47** (4), 1311-1317 (2014).
25. A. Dewaele, N. Worth, C. J. Pickard, R. J. Needs, S. Pascarelli, O. Mathon, M. Mezouar and T. Irifune, Nature Chemistry **8**, 784 (2016).
26. W. L. Vos, L. W. Finger, R. J. Hemley, J. Z. Hu, H. K. Mao and J. A. Schouten, Nature **358**, 46 (1992).
27. P. Loubeyre, M. Jean-Louis, R. LeToullec and L. Charon-Gérard, Physical Review Letters **70** (2), 178-181 (1993).
28. V. V. Struzhkin, D. Kim, E. Stavrou, T. Muramatsu, H. Mao, C. J. Pickard, R. J. Needs, V. B. Prakapenka and A. F. Goncharov, Nat Commun. **7**, 12267 (2016).
29. Z. M. Geballe, H. Liu, A. K. Mishra, M. Ahart, M. Somayazulu, Y. Meng, M. Baldini and R. J. Hemley, Angewandte Chemie International Edition **57** (3), 688-692 (2018).
30. C. Pépin, P. Loubeyre, F. Occelli and P. Dumas, Proceedings of the National Academy of Sciences of the United States of America **112** (25), 7673-7676 (2015).
31. C. M. Pépin, A. Dewaele, G. Geneste, P. Loubeyre and M. Mezouar, Physical Review Letters **113** (26), 265504 (2014).
32. D. Laniel, V. Svitlyk, G. Weck and P. Loubeyre, Physical Chemistry Chemical Physics **20** (6), 4050-4057 (2018).
33. D. K. Spaulding, G. Weck, P. Loubeyre, F. Datchi, P. Dumas and M. Hanfland, Nature Communications **5**, 5739 (2014).
34. A. F. Goncharov, N. Holtgrewe, G. Qian, C. Hu, A. R. Oganov, M. Somayazulu, E. Stavrou, C. J. Pickard, A. Berlie, F. Yen, M. Mahmood, S. S. Lobanov, Z. Konôpková and V. B. Prakapenka, The Journal of Chemical Physics **142** (21), 214308 (2015).
35. H. Wang, M. I. Eremets, I. Troyan, H. Liu, Y. Ma and L. Vereecken, Scientific Reports **5**, 13239 (2015).
36. P. Loubeyre, R. Le Toullec and J. P. Pinceaux, Physical Review B **36** (7), 3723-3730 (1987).
37. P. Loubeyre, R. Letoullec and J. P. Pinceaux, Journal of Physics: Condensed Matter **3** (18), 3183 (1991).
38. P. Loubeyre, R. LeToullec, D. Hausermann, M. Hanfland, R. J. Hemley, H. K. Mao and L. W. Finger, Nature **383** (6602), 702-704 (1996).
39. O. Pfaffenzeller, D. Hohl and P. Ballone, Physical Review Letters **74** (13), 2599-2602 (1995).
40. M. A. Morales, E. Schwegler, D. Ceperley, C. Pierleoni, S. Hamel and K. Caspersen, Proceedings of the National Academy of Sciences **106** (5), 1324-1329 (2009).
41. M. Schöttler and R. Redmer, Physical Review Letters **120** (11), 115703 (2018).
42. J. Vorberger, I. Tamblyn, B. Militzer and S. A. Bonev, Physical Review B **75** (2), 024206 (2007).
43. J. Lim and C.-S. Yoo, Physical Review Letters **120** (16), 165301 (2018).
44. R. Turnbull, M.-E. Donnelly, M. Wang, M. Peña-Alvarez, C. Ji, P. Dalladay-Simpson, H.-k. Mao, E. Gregoryanz and R. T. Howie, Physical Review Letters **121** (19),

195702 (2018).
45. A. F. Goncharov, S. S. Lobanov, V. B. Prakapenka and E. Greenberg, Physical Review B **95** (14), 140101 (2017).
46. H. K. Mao, J. Xu and P. M. Bell, Journal of Geophysical Research: Solid Earth **91** (B5), 4673-4676 (1986).
47. Y. Akahama, Y. Mizuki, S. Nakano, N. Hirao and Y. Ohishi, Journal of Physics: Conference Series **950** (4), 042060 (2017).
48. Y. Fei, A. Ricolleau, M. Frank, K. Mibe, G. Shen and V. Prakapenka, Proceedings of the National Academy of Sciences **104** (22), 9182-9186 (2007).
49. L. S. Dubrovinsky, S. K. Saxena and P. Lazor, Physics and Chemistry of Minerals **25** (6), 434-441 (1998).
50. See Supplemental Materials http://link.aps.org/supplemental/ for the detailed experimental data as obtained in this work, corresponding captions, and References.
51. P. Loubeyre, R. LeToullec and J. P. Pinceaux, Physical Review B **45** (22), 12844-12853 (1992).
52. Z. Liu, J. Botana, A. Hermann, S. Valdez, E. Zurek, D. Yan, H.-q. Lin and M.-s. Miao, Nature Communications **9** (1), 951 (2018).
53. R. J. Hemley, H. K. Mao and J. F. Shu, Physical Review Letters **65** (21), 2670-2673 (1990).
54. H.-k. Mao and R. J. Hemley, Reviews of Modern Physics **66** (2), 671-692 (1994).
55. R. T. Howie, I. B. Magdău, A. F. Goncharov, G. J. Ackland and E. Gregoryanz, Physical Review Letters **113** (17), 175501 (2014).
56. C. Ji, A. F. Goncharov, V. Shukla, N. K. Jena, D. Popov, B. Li, J. Wang, Y. Meng, V. B. Prakapenka, J. S. Smith, R. Ahuja, W. Yang and H.-k. Mao, Proceedings of the National Academy of Sciences **114** (14), 3596-3600 (2017).
57. H. K. Mao, R. J. Hemley, Y. Wu, A. P. Jephcoat, L. W. Finger, C. S. Zha and W. A. Bassett, Phys Rev Lett **60** (25), 2649-2652 (1988).
58. P. Loubeyre, R. LeToullec, J. P. Pinceaux, H. K. Mao, J. Hu and R. J. Hemley, Physical Review Letters **71** (14), 2272-2275 (1993).
59. A. Driessen, E. van der Poll and I. F. Silvera, Physical Review B **33** (5), 3269-3288 (1986).
60. M. S. Somayazulu, L. W. Finger, R. J. Hemley and H. K. Mao, Science **271** (5254), 1400-1402 (1996).
61. M. Somayazulu, P. Dera, A. F. Goncharov, S. A. Gramsch, P. Liermann, W. Yang, Z. Liu, H.-k. Mao and R. J. Hemley, Nature Chemistry **2**, 50 (2009).
62. P. Loubeyre, M. Jean-Louis, R. LeToullec and L. Charon-Gérard, Physical Review Letters **70** (2), 178-181 (1993).
63. S. Ninet, G. Weck, P. Loubeyre and F. Datchi, Physical Review B **83** (13), 134107 (2011).
64. Y. Akahama and H. Kawamura, Physical Review B **54** (22), R15602-R15605 (1996).
65. G. Weck, P. Loubeyre and R. LeToullec, Physical Review Letters **88** (3), 035504 (2002).

Supplemental Materials to

# Helium-hydrogen immiscibility at high pressures

Yu Wang[1,2], Xiao Zhang[1,2], Shuqing Jiang[1], Zachary M. Geballe[3], Teerachote Pakornchote[3,4], Maddury Somayazulu[3], Vitali Prakapenka[5], Eran Greenberg[5], Alexander F. Goncharov[1-3]

[1]*Key Laboratory of Materials Physics and Center for Energy Matter in Extreme Environments, Institute of Solid State Physics, Chinese Academy of Sciences, Hefei, Anhui 230031, China*
[2]*University of Science and Technology of China, Hefei 230026, Anhui, People's Republic of China*
[3]*Geophysical Laboratory, Carnegie Institution of Washington, 5251 Broad Branch Road, Washington, DC 20015, USA*
[4] *Department of Physics, Chulalongkorn University, Bangkok, 10330, Thailand*
[5]*Center for Advanced Radiations Sources, University of Chicago, Chicago, Illinois 60637, USA*

Alexander Goncharov
E-mail: agoncharov@carnegiescience.edu

This PDF file includes:
Figs. S1 to S5
References for SI reference citations

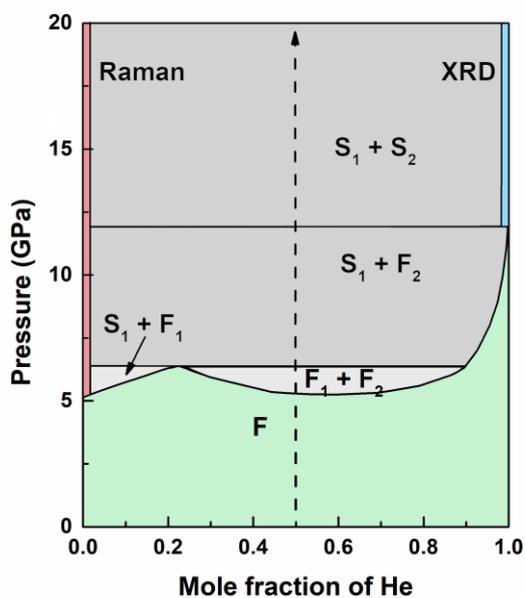

**Figure S1**. Binary phase diagram of $H_2$-He mixtures adapted from Loubeyre et al.[1]. Dashed line with arrow shows our experiment path. $F_1$, $S_1$, $F_2$ and $S_2$ signify fluid $H_2$, solid $H_2$, fluid He, and solid He, respectively. The areas colored green correspond to miscibility regions, while those color gray- are immiscible. Thick red and blue lines correspond to $H_2$ and He solids, observed in this study via Raman and XRD techniques, respectively.

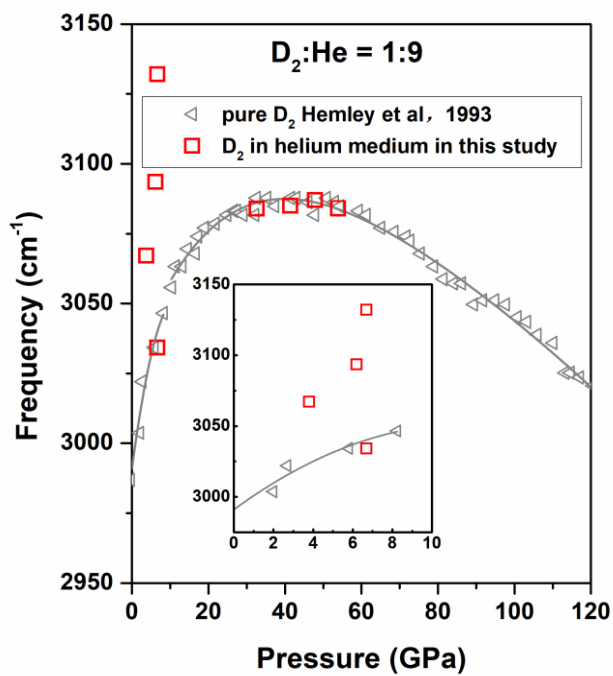

**Figure S2**. The pressure dependence of the vibron Raman frequency of the $D_2$-rich fluid and solid in He medium as a function of pressure. Red squares are the data of this study and gray triangles are the results of previous experiments in pure bulk $D_2$. [2]

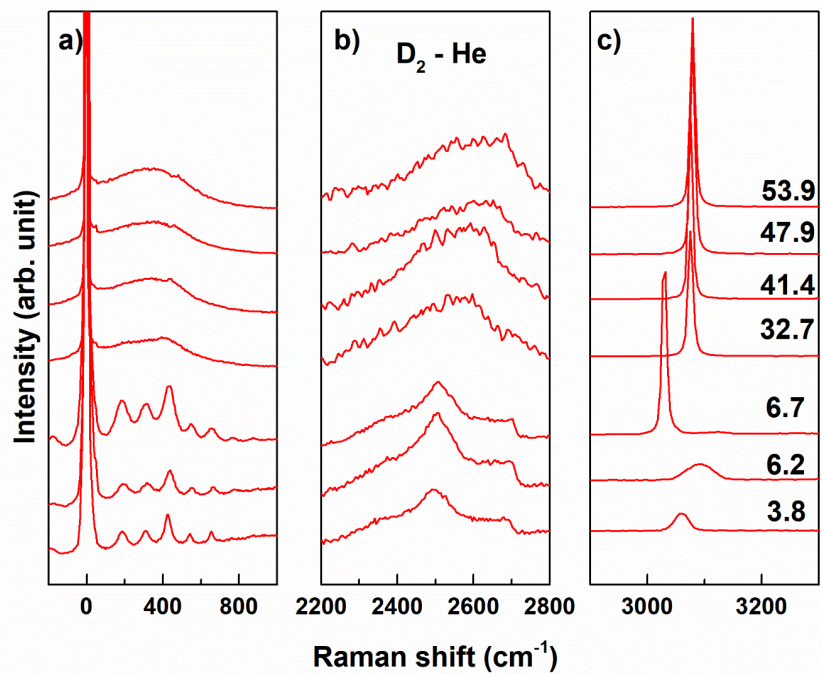

**Figure S3.** Raman spectra of $D_2$-He mixtures at different pressures in the range of rotational-translational (a), second-order diamond (b), and vibrational (c) modes.

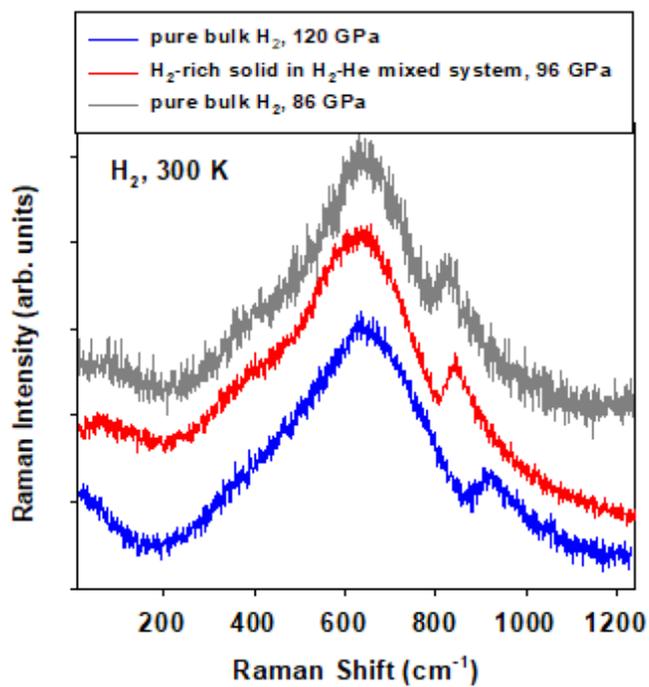

**Figure S4.** Raman spectra of $H_2$-rich solid formed in the mixed 50:50 $H_2$-He sample at 96 GPa compared to those of pure bulk $H_2$ at 86 and 120 GPa.

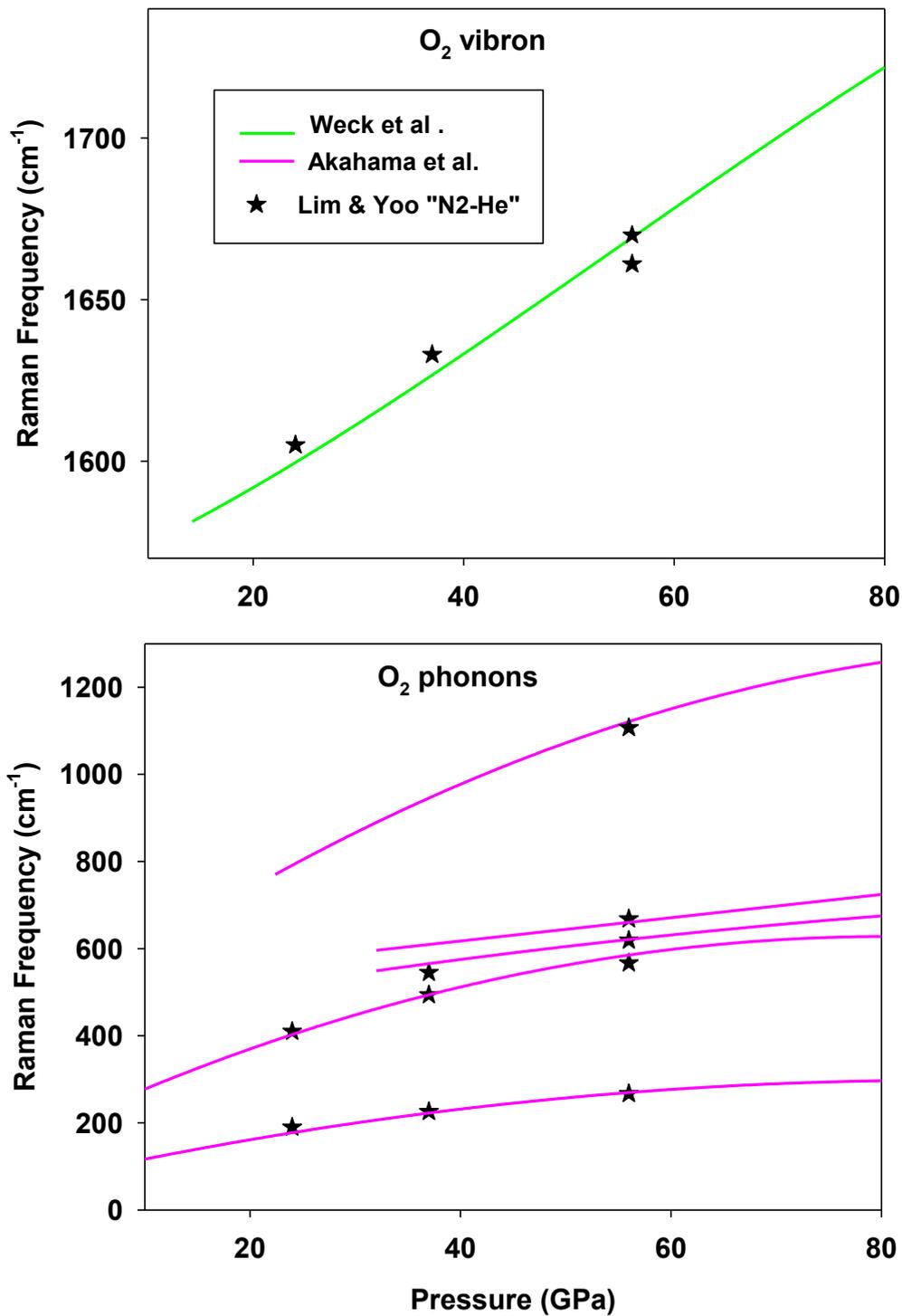

**Figure S5.** Pressure dependencies of the Raman frequencies of $O_2$ taken from the literature[3,4] compared to the reported by Lim and Yoo Raman frequencies of the $N_2$-He system[5]. The data suggest that Lim and Yoo investigated the contaminated by $O_2$ system.

**Supplemental References**

1. P. Loubeyre, R. Letoullec and J. P. Pinceaux, Journal of Physics: Condensed Matter **3** (18), 3183 (1991).
2. R. J. Hemley, J. H. Eggert and H.-k. Mao, Physical Review B **48** (9), 5779-5788 (1993).
3. G. Weck, P. Loubeyre and R. LeToullec, Physical Review Letters **88** (3), 035504 (2002).
4. Y. Akahama and H. Kawamura, Physical Review B **54** (22), R15602-R15605 (1996).
5. J. Lim and C.-S. Yoo, Physical Review Letters **120** (16), 165301 (2018).